\documentclass[aps,pre,twocolumn,superscriptaddress,showpacs]{revtex4-2}
\usepackage{graphicx} 

\usepackage{array}
\usepackage{amsfonts}
\usepackage{amssymb,amsmath,multirow,rotate}
\usepackage{bm}
\usepackage{color}
\usepackage{xr}
\usepackage{lipsum}

\definecolor{red}{rgb}{1,0,0}

\definecolor{yellow}{rgb}{1,0.5,0}
\definecolor{blue}{rgb}{0,0,1}
\definecolor{BLUE}{rgb}{0,0,1}
\definecolor{green}{rgb}{0,0.7,0}

\definecolor{lightgreen}{rgb}{0,0.85,0.35}

\definecolor{grey}{rgb}{0.6,0.6,0.6}

\definecolor{orange}{rgb}{1,0.6,0}

\begin{document}

\title{Basin Geometry and Reliable Recall of Dynamical Memories in Reservoir Computing}
\author{Ling-Wei Kong}  \email{Ling-Wei.Kong@cornell.edu}
\affiliation{Department of Computational Biology, Cornell University, Ithaca, NY 14850, USA}
\author{Ying-Cheng Lai} 
\affiliation{School of Electrical, Computer, and Energy Engineering, Arizona State University, Tempe, AZ 85287, USA}
\affiliation{Department of Physics, Arizona State University, Tempe, Arizona 85287, USA}
\date{\today}

\begin{abstract}

Reliable attractor recall conventionally requires broad basins of attraction. However, in reservoir-computing based 
associative memory, temporal cues reliably recover dynamical memories despite basins dominated by unpredictable, riddled-like regions. We reveal that memory basins exhibit an ``octopus-like'' structure: a robust ``head'' near the attractor and thin, intertwined ``tentacles'' spanning state space. Initial states in tentacular regions yield near-zero uncertainty exponents, 
making the recalled memory effectively unpredictable at finite precision.
Yet, cue-driven generalized synchronization bypasses this unpredictability, driving the system into the robust basin head. This mechanism yields a quantitative relation linking minimum cue duration, synchronization rate, and basin-head radius. Trained recurrent neural networks exhibit similar geometry, suggesting this phenomenon extends beyond reservoir computing.


\end{abstract}

\maketitle

Does reliable recall require a geometrically robust basin of attraction? The classical attractor picture of associative memory 
says yes: a partial or noisy cue initializes the network state, the recurrent dynamics then run autonomously to a stored attractor, and the basin fixes how much cue error can be corrected.
Large, coherent basins should support reliable recall. This intuition underlies both classical Hopfield-type memories and the common use of basin geometry or basin probability as a measure of recall robustness~\cite{Hopfield:1982,amit1985spin,HT:1986,storkey1999basins,menck2013basin,rubin2017balanced,marsh2021enhancing}. The same picture extends to dynamical memories such as sequences and oscillations~\cite{sompolinsky1986temporal,pals2024trained,pereira2023forgetting,essex2026memorization}, accessed from a short cue segment. Yet once the cue is removed, the network must continue autonomously within the target basin, so the classical requirement seems to persist. 

\begin{figure*}[ht]
\centering
\includegraphics[width=\linewidth]{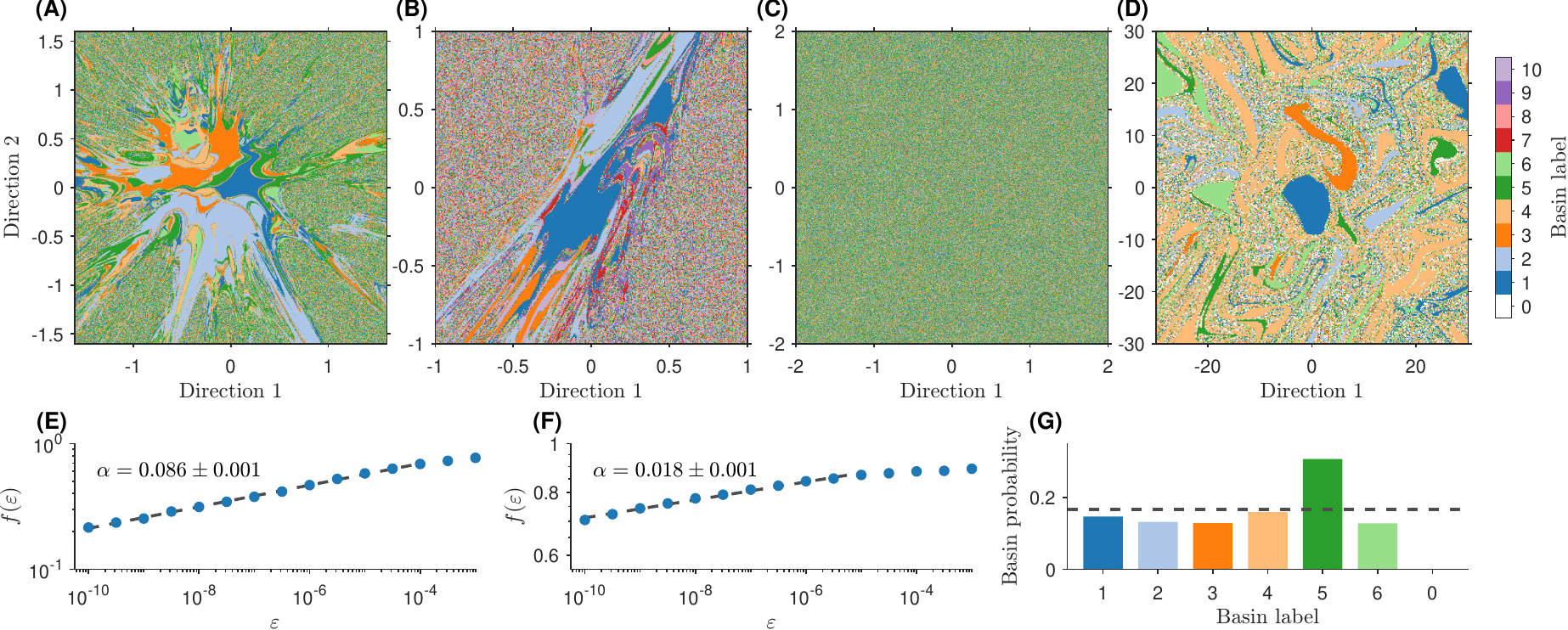}
\caption{\textbf{Octopus-like basins with riddled-like tentacles.} Basin geometry is visualized through two-dimensional slices of the high-dimensional recurrent state space. (A,B) Two-dimensional basin slices for reservoirs storing six and ten dynamical memories, respectively. Each slice is centered on a randomly selected point of the memory-1 trajectory and spanned by two random orthonormal directions. A locally robust head surrounds the trajectory, while thin, interwoven tentacles extend across the slice. (C) Slice centered on a random reservoir state, with nearly the entire displayed region riddled-like. (D) A recurrent neural network trained by backpropagation with low-rank adaptation (LoRA) shows similar head-and-tentacle geometry. 
(E,F) Uncertain fraction $f(\varepsilon)$ versus perturbation scale $\varepsilon$ for the six- and ten-memory reservoirs. Fits to $f(\varepsilon)\sim\varepsilon^\alpha$ yield $\alpha=0.086$ and $0.018$, respectively. (G) Basin probabilities estimated from 1,200 random initial conditions for the six-memory reservoir. All trials reached one of the six learned memories, each attracting a fraction of order 1/6 of the sampled initial conditions, though not equally.
}
\label{fig:basin}
\end{figure*}

We encounter a strikingly different situation in the reservoir-computing memory system introduced in our previous work, where multiple dynamical attractors are recalled reliably once a temporal cue exceeds a short characteristic duration~\cite{kong2024reservoir}. A reservoir computer is a high-dimensional recurrent network with typically fixed internal connections and a trained linear readout~\cite{Jaeger:2001,MNM:2002,JH:2004}. With its trained output fed back as input, the same network becomes autonomous. Previous studies have used reservoirs to realize multiple learned dynamics, including coexisting attractors~\cite{inoue2020designing,ceni2020interpreting,lu2020invertible,kong2021machine,kim2021teaching,flynn2021multifunctionality,flynn2022exploring,KB:2023,kong2024reservoir}. 
Examining the basin structure of multiple memory states, as in Ref.~\cite{kong2024reservoir}, we find an extremely complicated geometry. The basins exhibit the octopus-like geometry introduced by Zhang and Strogatz~\cite{zhang2021basins} in coupled Kuramoto oscillators, each with a locally robust ``head'' around the attractor and thin ``tentacles'' that thread through those of other memories. Moreover, the tentacular regions in our reservoir computers form a riddled-like region across much of the sampled state space (Fig.~\ref{fig:basin}A--C). 
An exact state in these riddled-like regions still has a unique asymptotic destination, yet its finite-resolution neighborhood can contain states leading to many different memories. Shrinking perturbations by five orders of magnitude barely restores predictability, yielding uncertainty exponents close to zero. Basin size or probability can therefore no longer track recoverability. How can recall remain reliable when the basin landscape offers so little local certainty?

Unlike an instantaneous initialization, a temporal cue continues to act on the network. Its temporal structure directs the reservoir toward the target memory, while repeated driving washes out its dependence on the initial state. Together, these effects can be understood as generalized synchronization between the cue and the reservoir response~\cite{rulkov1995generalized,kocarev1996generalized,hunt1997differentiable,lu2018attractor,lymburn2019consistency,kim2025sigma}. What matters is then not where the network starts but how much uncertainty remains when autonomous dynamics take over. The recall threshold becomes a matching problem between the rate at which cueing removes uncertainty and the amount that the target basin can tolerate, which we turn into a predictive, parameter-free law for the critical cue length.

The answer lies in the robust basin head, not in the tentacles. We show that recall becomes reliable once the residual state uncertainty fits inside the head, even though the surrounding riddled-like geometry remains unchanged. The relevant geometric scale is therefore the head radius rather than the total basin volume or the reach of its tentacles.


We first examine the autonomous basin geometry of the learned dynamical memories. We focus on the memories of periodic states in the main text, while analogous results for chaotic memories are presented in the Supplemental Material. Because the reservoir state space is high-dimensional, the full basins cannot be visualized directly. We therefore examine basin structure through two-dimensional slices of the reservoir state space. The origin of each slice is a randomly selected point on the trajectory of memory 1, and the slice is spanned by two random orthonormal directions. Each point on the slice is used as an initial condition and labeled by the attractor reached under autonomous evolution. Figures~\ref{fig:basin}A and \ref{fig:basin}B show representative slices for reservoirs storing six and ten memories. The same qualitative organization appears in slices with different origins, orientations, and reservoir computers (Supplemental Material).

The basin geometry is strikingly complex. Across the sampled slices, the basins exhibit an octopus-like organization~\cite{zhang2021basins}. A locally robust head surrounds the memory trajectory, while thin tentacles extend far into the surrounding state space and interweave with those of other memories. Within the head, small perturbations usually leave the final attractor unchanged. A point in a tentacle still has a well-defined asymptotic destination, but a nearby point can converge to a different memory. Asymptotic basin membership is therefore unambiguous, even where finite-precision recoverability is extremely poor.

The fine-scale intermingling of the tentacles resembles that of riddled basins and intermingled basins~\cite{alexander1992riddled,ott1994transition,kan1994open,lai1995intermingled}, although we do not claim that the basins satisfy the strict mathematical definition of either. We therefore refer to these regions as riddled-like. To quantify how outcome uncertainty changes with scale, we measure the uncertainty exponent. For a perturbation scale $\varepsilon$, let $f(\varepsilon)$ denote the fraction of initial conditions whose final attractor changes after a perturbation of magnitude $\varepsilon$. The scaling \begin{align} 
f(\varepsilon)\sim \varepsilon^\alpha 
\end{align} 
defines the uncertainty exponent $\alpha$. In a two-dimensional slice, a smooth basin boundary gives $\alpha=1$, while increasingly fractal boundaries give smaller values of $\alpha$. As $\alpha$ approaches zero, reducing the perturbation scale produces almost no reduction in uncertain outcomes. As shown in Figs.~\ref{fig:basin}E and \ref{fig:basin}F, we obtain $\alpha=0.086$ for the reservoir storing six memories and $\alpha=0.018$ for the reservoir storing ten memories. For the ten-memory reservoir, reducing the perturbation scale by five orders of magnitude lowers the uncertain fraction by less than 0.1 order of magnitude. Thus, a 100,000-fold improvement in state precision yields little improvement in predicting the final attractor. These near-zero exponents provide quantitative support for describing the tentacular regions as riddled-like. For recall, this means that placing the reservoir somewhere in the correct mathematical basin is not enough. If the cue is removed while the state remains in a tentacular region, extremely small residual errors can redirect the autonomous dynamics to another memory. Reliable recall requires the cue to carry the state into the robust head of the target basin, whose effective margin is quantified below by $r_{\rm head}$.

The slices in Figs.~\ref{fig:basin}A and \ref{fig:basin}B are deliberately centered on a stored trajectory, making the robust head visually prominent. To determine whether the intermingled geometry is confined to its vicinity, we also examine slices centered on randomly selected reservoir states. Figure~\ref{fig:basin}C shows a representative example in which almost the entire displayed region is riddled-like, indicating that the tentacular organization extends far beyond the visible heads. Yet small heads do not imply small basins. To estimate the basin probabilities, we evolved 1,200 randomly sampled initial states under autonomous dynamics [Fig.~\ref{fig:basin}G]. Every trial converged to one of the six learned memories, and no other asymptotic outcome was observed. Each memory was reached from a fraction of initial states of order $1/6$. 
Under this sampling measure, every memory basin therefore occupies a substantial fraction of the sampled state space. Together with the localized heads seen in the slices, these basin probabilities suggest that most of the basin measure lies in the extended tentacular regions. 

A qualitatively similar head-and-tentacle organization appears in a recurrent neural network trained by backpropagation with low-rank adaptation [Fig.~\ref{fig:basin}D]. This single example does not establish universality across recurrent architectures, but it shows that the coexistence of robust heads and intermingled tentacles is not tied to the fixed random recurrent matrix used in conventional reservoir computing. (More details in the Supplemental Material.)

This geometry creates an apparent puzzle. How can a finite cue produce reliable recall when autonomous basin outcomes are so sensitive to uncertainty? A temporal cue does not require the reservoir to begin in the correct autonomous basin. It changes the dynamics, drives the state across the autonomous basin landscape, and progressively suppresses its dependence on the initial condition. We show next that this process is governed by generalized synchronization and that reliable recall occurs once the driven state reaches the robust head of the target basin.

\begin{figure}[htb]
\centering
\includegraphics[width=0.96\linewidth]{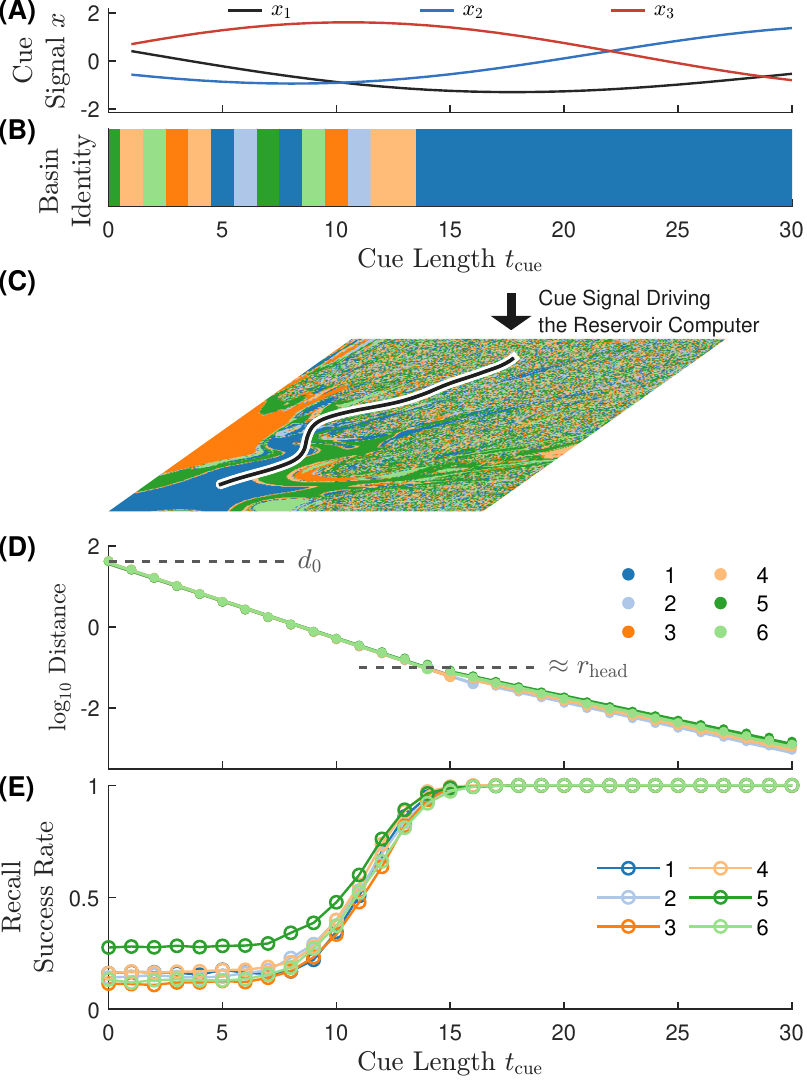}
\caption{\textbf{Cue-driven contraction into a robust basin head.} (A) Example three-component temporal cue. (B) Basin identity of the reservoir state after cue prefixes of increasing length. At each cue length, the cue is removed and the state is evolved autonomously; colors indicate the resulting attractor. Short cues leave the state in different basins, whereas longer cues place it consistently in the target basin. (C) Schematic plot of the cue-driven state contracting through the intermingled basin region into the robust head of the target basin. (D) For cues corresponding to all six memories, the distance $d(t)$ between the driven state and the target memory trajectory decreases approximately exponentially from the typical initial scale $d_0$. The lower dashed line marks the basin-head scale $r_{\rm head}$. 
(E) Recall success rate over random initial conditions rises sharply as $d(t)$ approaches $r_{\rm head}$. Numbers and colors identify the six target memories.}
\label{fig:sync}
\end{figure}


During recall, the reservoir is driven by a short segment of the target memory [Fig.~\ref{fig:sync}A]. To determine where the driven state lies relative to the autonomous basins, we remove the cue after successive cue lengths and evolve the resulting state autonomously. For short cues, the basin identity changes repeatedly as the driven state crosses the intermingled basins [Fig.~\ref{fig:sync}B]. After sufficient driving, the state enters and remains in the target basin. 

The minimum cue length needed to reach a prescribed recall probability provides an operational measure of recall robustness. In a Hopfield-type associative memory, robustness is reflected by the ability to recover a stored pattern from an incomplete or noisy static cue. The temporal analog is whether a short segment of a dynamical memory contains enough information to bring the reservoir into the robust head of its target basin, from which recall can proceed autonomously. A shorter critical cue length therefore means that less of the target memory must be supplied for reliable access.
To measure the underlying contraction, we apply an adapted version of the auxiliary-system approach~\cite{abarbanel1996generalized}. We first drive a reference copy of the reservoir from a random initial state with the target cue for $T_{\rm prep}=300$ steps, well beyond the observed recall threshold. By then, the reference copy has reached generalized synchronization with the cue, and its subsequent hidden state represents the cue-synchronized trajectory associated with the target memory~\cite{rulkov1995generalized,kocarev1996generalized,hunt1997differentiable,lu2018attractor,lymburn2019consistency}. We then initialize a second copy of the same reservoir from an independent random state and drive both copies with the same continuation of the cue. Setting $t=0$ at the beginning of this continuation, we denote the reference trajectory by $\mathbf{r}_{}(t)$ and the newly initialized trajectory by $\mathbf{r}(t;\mathbf{r}_0)$. Their separation is
\begin{align}
d(t)=\lVert\mathbf{r}(t;\mathbf{r}0)-\mathbf{r}{}(t)\rVert_2 .
\end{align}
The decay of $d(t)$ measures how rapidly the cue removes the influence of the new initial condition and brings the reservoir toward the target memory trajectory.

For all six memory cues, this distance decreases approximately exponentially over the contraction regime [Fig.~\ref{fig:sync}D],
\begin{align}
d(t)\approx d_0\exp\left(-\lambda_{\rm sync}t\right), \label{eq:d_exp}
\end{align}
where $d_0$ is the typical separation at $t=0$ and $\lambda_{\rm sync}>0$ is the finite-time synchronization rate.
Given the high dimensionality of the reservoir state space, we can calculate $d_0$ directly by $d_0\approx\sqrt{2N/3}$ derived in the Supplemental Material.
This contraction describes a finite-time approach toward generalized synchronization. Complete synchronization is unnecessary because recall can proceed once the driven state reaches the robust head of the target basin. We denote by $r_{\rm head}$ the effective perturbation scale around the target trajectory below which autonomous evolution reliably returns to the target memory. The critical cue length is therefore determined by
\begin{align}
d(t_{\rm recall})\approx r_{\rm head}.
\end{align}
The recall success rate rises sharply near this crossing, as shown in Fig.~\ref{fig:sync}E. In the present system, a cue segment shorter than $1/20$ of the corresponding memory period is sufficient for reliable recall. Thus, the reservoir can recover the target memory after observing only a small fraction of its temporal pattern.


\begin{figure}
\centering
\includegraphics[width=0.75\linewidth]{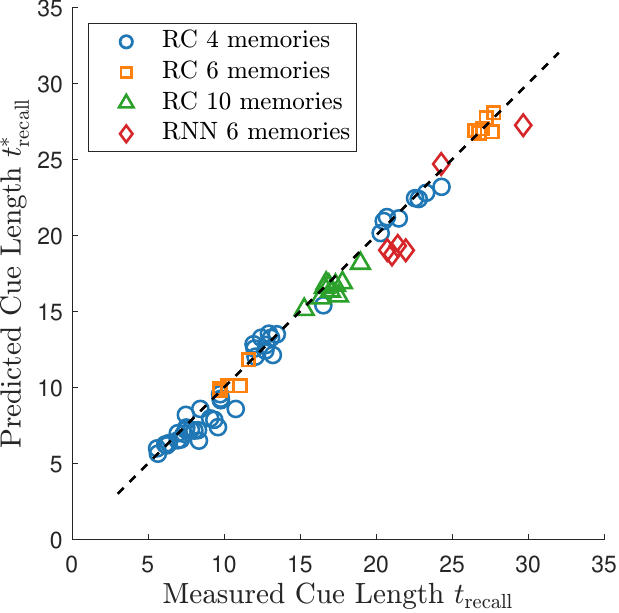}
\caption{\textbf{Geometry--dynamics prediction of memory recall time.} Predicted versus measured critical cue duration $t_{\rm recall}$ for reservoir networks and RNNs spanning different hyperparameters and memory states (Supplemental Table~1). Each point represents one stored memory in one network. Predictions are obtained from Eq.~\eqref{eq:recall} using independently measured values of $\lambda_{\rm sync}$ and $r_{\rm head}$. The measured $t_{\rm recall}$ is the shortest cue duration for which at least $90\%$ of random initial conditions recall the target memory. The diagonal line denotes equality between prediction and measurement. The measured recall times closely follow the predictions obtained from independently measured quantities, with a mean absolute error of 0.65 time steps, and a predictive $R^2$ of 0.98.}
\label{fig:eq}
\end{figure}

The critical cue duration is therefore set by the crossing of the cue-driven distance and the autonomous tolerance, $d(t_{\rm recall})\approx r_{\rm head}$. Substituting Eq.~\eqref{eq:d_exp} gives
\begin{align}
t_{\rm recall}
\approx
\frac{1}{\lambda_{\rm sync}}
\log\left(\frac{d_0}{r_{\rm head}}\right).
\label{eq:recall}
\end{align}
Equation~\eqref{eq:recall} connects the cue-driven dynamics, through the synchronization rate $\lambda_{\rm sync}$, with the autonomous basin geometry, through the robust head radius $r_{\rm head}$.
A smaller basin head requires a longer cue, whereas faster cue-induced contraction shortens the cue needed to reach the same head.
Within a given reservoir network, we find that $\lambda_{\rm sync}$ varies little among the stored memories, suggesting that it is primarily a dynamical property of the network.
The basin geometry, including $r_{\rm head}$, remains memory-dependent.
The law therefore separates a network-level contraction rate from a memory-specific geometric tolerance.

We define $r_{\rm head}$ and $t_{\rm recall}$ using the same reliability threshold $p$. To measure $r_{\rm head}$, we sample states along the target trajectory, perturb them over random directions, and identify the perturbation scale at which the probability of autonomous return to the target memory falls below $p$. We define $t_{\rm recall}$ as the shortest cue duration for which recall succeeds from at least a fraction $p$ of random initial states.
We use $p=0.9$ throughout the main text. Alternative thresholds give the same qualitative results (Supplemental Material).
We test Eq.~\eqref{eq:recall} across reservoir networks spanning a range of hyperparameters, and also on LoRA recurrent neural networks. For each network and memory, $\lambda_{\rm sync}$ and $r_{\rm head}$ are measured independently, without fitting the observed recall time. The measured recall times closely follow the parameter-free predictions of Eq.~\eqref{eq:recall} [Fig.~\ref{fig:eq}].


Our results separate two notions that are often implicitly conflated in attractor-based descriptions of memory: asymptotic basin membership and reliable accessibility at finite precision. States on the thin basin tentacles belong mathematically to well-defined attractor basins, yet small perturbations can redirect nearby states toward different memories. The coherent heads and intermingled tentacles therefore play qualitatively different roles despite belonging to the same basins. 
Because a memory can occupy a large share of state space while its robust head stays small (Fig.~\ref{fig:basin}G), total basin volume need not provide a useful measure of memory robustness. It does not distinguish locally stable regions from regions lying arbitrarily close to competing basins. For finite-precision recall, the more relevant geometric quantity is the robust margin $r_{\rm head}$ around the state reached during cueing.

The apparent tension between reliable recall and strongly intermingled autonomous basins disappears once the cue-driven dynamics is taken into account. During cueing, the reservoir evolves under a driven system, and its trajectory is not constrained to remain within one autonomous basin. Instead, the cue suppresses dependence on the initial state through generalized synchronization. Once the cue is removed, the resulting state is captured by whichever autonomous basin it has reached. Memory retrieval can therefore be viewed as a finite-time synchronization process with a geometrically determined stopping condition, rather than as navigation along the correct autonomous basin tentacle.

This picture also provides a dynamical analogue of pattern completion in conventional associative memory. In our examples, a cue segment shorter than one twentieth of the corresponding memory period can be sufficient for reliable recall. Thus, only a small fraction of the temporal pattern is needed to deliver an initially unknown reservoir state into the robust head of the target basin, much as a partial or noisy static pattern can retrieve a stored state in a Hopfield network. The geometry--dynamics relation Eq.~\eqref{eq:recall} makes the two contributions explicit. The synchronization rate $\lambda_{\rm sync}$ controls how rapidly the cue removes initial-state uncertainty, whereas $r_{\rm head}$ determines how much residual uncertainty the autonomous memory can tolerate.


Equation~\eqref{eq:recall} suggests that recall may be improved by strengthening cue-induced contraction or enlarging robust heads. Future work could test whether these properties can be tuned independently and how $r_{\rm head}$ and $t_{\rm recall}$ scale with memory load, including a possible capacity--accessibility tradeoff. The RNN result motivates testing whether the same threshold law applies across recurrent architectures and with incomplete or noisy cues. 

L.-W. K. acknowledges support from the Eric and Wendy Schmidt AI in Science Postdoctoral Fellowship, a program of Schmidt Sciences, LLC. The work at Arizona State University was supported by the US Army Research Office under Grants No. W911NF-26-2-A002 and No. W911NF-24-2-0228.



\bibliography{memory,RC_basin}

@article{kong2024reservoir,
  title={Reservoir-computing based associative memory and itinerancy for complex dynamical attractors},
  author={Kong, Ling-Wei and Brewer, Gene A and Lai, Ying-Cheng},
  journal={Nat. Commun.},
  volume={15},
  number={1},
  pages={4840},
  year={2024},
  publisher={Nature Publishing Group UK London}
}

@article{alexander1992riddled,
  title={Riddled basins},
  author={Alexander, JC and Yorke, James A and You, Zhiping and Kan, Ittai},
  journal={Int. J. Bif. Chaos},
  volume={2},
  number={04},
  pages={795--813},
  year={1992},
  publisher={World Scientific}
}

@article{ott1994transition,
  title={The transition to chaotic attractors with riddled basins},
  author={Ott, Edward and Alexander, JC and Kan, Ittai and Sommerer, John C and Yorke, James A},
  journal={Physica D},
  volume={76},
  number={4},
  pages={384--410},
  year={1994},
  publisher={Elsevier}
}

@article{kan1994open,
  title={Open sets of di feomorphisms having two attractors each with an everywhere dense basin},
  author={Kan, Ittai},
  journal={Bull. Ame. Math. Soc.},
  volume={31},
  number={1},
  year={1994}
}

@article{lai1995intermingled,
  title={Intermingled basins and two-state on-off intermittency},
  author={Lai, Ying-Cheng and Grebogi, Celso},
  journal={Phys. Rev. E},
  volume={52},
  number={4},
  pages={R3313},
  year={1995},
  publisher={APS}
}

@article{Hopfield:1982,
  title={Neural networks and physical systems with emergent collective computational abilities.},
  author={Hopfield, John J},
  journal={Proc. Nat. Acad. Sci. (USA)},
  volume={79},
  number={8},
  pages={2554--2558},
  year={1982},
  publisher={National Acad Sciences}
}

@article{amit1985spin,
  title={Spin-glass models of neural networks},
  author={Amit, Daniel J and Gutfreund, Hanoch and Sompolinsky, Haim},
  journal={Phys. Rev. A},
  volume={32},
  number={2},
  pages={1007},
  year={1985},
  publisher={APS}
}

@article{HT:1986,
  title={Computing with neural circuits: A model},
  author={Hopfield, John J and Tank, David W},
  journal={Science},
  volume={233},
  number={4764},
  pages={625--633},
  year={1986},
  publisher={American Association for the Advancement of Science}
}

@article{essex2026memorization,
  title={Memorization and forgetting in a learning {Hopfield neural} network: Bifurcation mechanisms, attractors, and basins},
  author={Essex, Adam E and Janson, Natalia B and Norris, Rachel A and Balanov, Alexander G},
  journal={Chaos},
  volume={36},
  number={8},
  year={2026},
  publisher={AIP Publishing}
}

@article{pals2024trained,
  title={Trained recurrent neural networks develop phase-locked limit cycles in a working memory task},
  author={Pals, Matthijs and Macke, Jakob H and Barak, Omri},
  journal={PLOS Comp. Biol.},
  volume={20},
  number={2},
  pages={e1011852},
  year={2024},
  publisher={Public Library of Science San Francisco, CA USA}
}

@article{menck2013basin,
  title={How basin stability complements the linear-stability paradigm},
  author={Menck, Peter J and Heitzig, Jobst and Marwan, Norbert and Kurths, J{\"u}rgen},
  journal={Nat. Phys.},
  volume={9},
  number={2},
  pages={89--92},
  year={2013},
  publisher={Nature Publishing Group UK London}
}

@article{storkey1999basins,
  title={The basins of attraction of a new {Hopfield} learning rule},
  author={Storkey, Amos J and Valabregue, Romain},
  journal={Neu. Net.},
  volume={12},
  number={6},
  pages={869--876},
  year={1999},
  publisher={Elsevier}
}

@article{pereira2023forgetting,
  title={Forgetting leads to chaos in attractor networks},
  author={Pereira-Obilinovic, Ulises and Aljadeff, Johnatan and Brunel, Nicolas},
  journal={Phys. Rev. X},
  volume={13},
  number={1},
  pages={011009},
  year={2023},
  publisher={APS}
}

@article{sompolinsky1986temporal,
  title={Temporal association in asymmetric neural networks},
  author={Sompolinsky, Haim and Kanter, Ido},
  journal={Phys. Rev. Lett.},
  volume={57},
  number={22},
  pages={2861},
  year={1986},
  publisher={APS}
}

@article{marsh2021enhancing,
  title={Enhancing associative memory recall and storage capacity using confocal cavity QED},
  author={Marsh, Brendan P and Guo, Yudan and Kroeze, Ronen M and Gopalakrishnan, Sarang and Ganguli, Surya and Keeling, Jonathan and Lev, Benjamin L},
  journal={Phys. Rev. X},
  volume={11},
  number={2},
  pages={021048},
  year={2021},
  publisher={APS}
}

@article{rubin2017balanced,
  title={Balanced excitation and inhibition are required for high-capacity, noise-robust neuronal selectivity},
  author={Rubin, Ran and Abbott, Larry F and Sompolinsky, Haim},
  journal={Proc. Nat. Acad. Sci.},
  volume={114},
  number={44},
  pages={E9366--E9375},
  year={2017},
  publisher={National Academy of Sciences}
}

@article{kocarev1996generalized,
  title={Generalized synchronization, predictability, and equivalence of unidirectionally coupled dynamical systems},
  author={Kocarev, Ljupco and Parlitz, Ulrich},
  journal={Phys. Rev. Lett.},
  volume={76},
  number={11},
  pages={1816},
  year={1996},
  publisher={APS}
}

@article{hunt1997differentiable,
  title={Differentiable generalized synchronization of chaos},
  author={Hunt, Brian R and Ott, Edward and Yorke, James A},
  journal={Phys. Rev. E},
  volume={55},
  number={4},
  pages={4029},
  year={1997},
  publisher={APS}
}

@article{abarbanel1996generalized,
  title={Generalized synchronization of chaos: The auxiliary system approach},
  author={Abarbanel, Henry DI and Rulkov, Nikolai F and Sushchik, Mikhail M},
  journal={Phys. Rev. E},
  volume={53},
  number={5},
  pages={4528},
  year={1996},
  publisher={APS}
}

@article{lu2018attractor,
  title={Attractor reconstruction by machine learning},
  author={Lu, Zhixin and Hunt, Brian R and Ott, Edward},
  journal={Chaos},
  volume={28},
  number={6},
  year={2018},
  publisher={AIP Publishing}
}

@article{lymburn2019consistency,
  title={Consistency in echo-state networks},
  author={Lymburn, Thomas and Khor, Alexander and Stemler, Thomas and Corr{\^e}a, D{\'e}bora C and Small, Michael and J{\"u}ngling, Thomas},
  journal={Chaos},
  volume={29},
  number={2},
  year={2019},
  publisher={AIP Publishing}
}

@article{kim2025sigma,
  title={{SIGMa-DS}: System identification from the geometric manifold of dynamical synchronization},
  author={Kim, Jason Z and Kong, Ling-Wei and Lu, Zhixin},
  journal={Chaos},
  volume={35},
  number={10},
  year={2025},
  publisher={AIP Publishing}
}

@article{Jaeger:2001,
title={The "echo state" approach to analysing and training recurrent neural networks-with an erratum note},
author={Jaeger, Herbert},
journal={Bonn, Germany: German National Research Center for Information Technology GMD Technical Report},
volume={148},
number={34},
pages={13},
year={2001},
publisher={Bonn}
}

@article{rulkov1995generalized,
  title = {Generalized synchronization of chaos in directionally coupled chaotic systems},
  author = {Rulkov, Nikolai F. and Sushchik, Mikhail M. and Tsimring, Lev S. and Abarbanel, Henry D. I.},
  journal = {Phys. Rev. E},
  volume = {51},
  issue = {2},
  pages = {980--994},
  numpages = {0},
  year = {1995},
  month = {Feb},
  publisher = {American Physical Society},
  doi = {10.1103/PhysRevE.51.980}
}

@article{lu2020invertible,
  title={Invertible generalized synchronization: A putative mechanism for implicit learning in neural systems},
  author={Lu, Zhixin and Bassett, Danielle S},
  journal={Chaos},
  volume={30},
  number={6},
  pages = {063133},
  year={2020},
  publisher={AIP Publishing}
}

@article{flynn2021multifunctionality,
  title={Multifunctionality in a reservoir computer},
  author={Flynn, Andrew and Tsachouridis, Vassilios A and Amann, Andreas},
  journal={Chaos},
  volume={31},
  number={1},
  pages = {013125}, 
  year={2021},
  publisher={AIP Publishing}
}

@inproceedings{flynn2022exploring,
  title={Exploring the limits of multifunctionality across different reservoir computers},
  author={Flynn, Andrew and Heilmann, Oliver and K{\"o}glmayr, Daniel and Tsachouridis, Vassilios A and R{\"a}th, Christoph and Amann, Andreas},
  booktitle={2022 International Joint Conference on Neural Networks (IJCNN)},
  pages={1--8},
  year={2022},
  organization={IEEE}
}

@article{zhang2021basins,
  title = {Basins with Tentacles},
  author = {Zhang, Yuanzhao and Strogatz, Steven H.},
  journal = {Phys. Rev. Lett.},
  volume = {127},
  issue = {19},
  pages = {194101},
  numpages = {6},
  year = {2021},
  month = {Nov},
  publisher = {American Physical Society},
  doi = {10.1103/PhysRevLett.127.194101}
}

@article{kong2021machine,
  title={Machine learning prediction of critical transition and system collapse},
  author={Kong, Ling-Wei and Fan, Hua-Wei and Grebogi, Celso and Lai, Ying-Cheng},
  journal={Phys. Rev. Res.},
  volume={3},
  number={1},
  pages={013090},
  year={2021},
  publisher={APS}
}

@article{kim2021teaching,
  title={Teaching recurrent neural networks to infer global temporal structure from local examples},
  author={Kim, Jason Z and Lu, Zhixin and Nozari, Erfan and Pappas, George J and Bassett, Danielle S},
  journal={Nat. Mach. Intell.},
  volume={3},
  number={4},
  pages={316--323},
  year={2021},
  publisher={Nature Publishing Group UK London}
}

@article{JH:2004,
  title = {Harnessing nonlinearity: Predicting chaotic systems and saving energy in wireless communication},
  author = {H. Jaeger and H. Haas},
  journal = {Science},
  volume = {304},
  pages = {78-80},
  year = {2004}
}

@article{MNM:2002,
  title={Real-time computing without stable states: A new framework for neural computation based on perturbations},
  author={Maass, Wolfgang and Natschl{\"a}ger, Thomas and Markram, Henry},
  journal={Neur. Comp.},
  volume={14},
  number={11},
  pages={2531--2560},
  year={2002},
  publisher={MIT Press}
}

@article{ceni2020interpreting,
  title={Interpreting recurrent neural networks behaviour via excitable network attractors},
  author={Ceni, Andrea and Ashwin, Peter and Livi, Lorenzo},
  journal={Cogn. Comp.},
  volume={12},
  pages={330--356},
  year={2020},
  publisher={Springer}
}

@article{inoue2020designing,
  title={Designing spontaneous behavioral switching via chaotic itinerancy},
  author={Inoue, Katsuma and Nakajima, Kohei and Kuniyoshi, Yasuo},
  journal={Sci. Adv.},
  volume={6},
  number={46},
  pages={eabb3989},
  year={2020},
  publisher={American Association for the Advancement of Science}
}

@article{KB:2023,
  title={A neural machine code and programming framework for the reservoir computer},
  author={Kim, Jason Z. and Bassett, Dani S.}, 
  journal={Nat. Mach. Intell.},
  volume={5},
  pages={622-630},
  year={2023}
}

\end{document}